\documentclass[12pt]{article}

\usepackage{float,color}
\usepackage{graphicx}
\usepackage{times}
\usepackage{amsmath,amssymb}
\usepackage{url}
\usepackage{subfig}
\usepackage[left,displaymath,mathlines]{lineno}

\newcounter{lastnote}

\title{Urban Scaling in Europe}

\author
{Lu\'is M. A. Bettencourt$^{1,*}$, Jos\'e Lobo$^2$ \\
\normalsize{$^1$Santa Fe Institute, 1399 Hyde Park Rd, Santa Fe NM 87501, USA,}
\\
\normalsize{$^{2}$ School of Sustainability, Arizona State University, 800 Cady Mall, Tempe, AZ 85281, USA.}
\\
\normalsize{$^*$To whom correspondence should be addressed; E-mail:  bettencourt@santafe.edu}
}
\date{\today}

\begin{document}

\baselineskip24pt

\maketitle


%\begin{sciabstract}
\begin{abstract}
Over the last decades, in disciplines as diverse as economics, geography, and complex systems, a perspective has arisen proposing that many properties of cities are quantitatively predictable due to agglomeration or scaling effects. Using new harmonized definitions for functional urban areas, we examine to what extent these ideas apply to European cities. We show that while most large urban systems in Western Europe (France, Germany, Italy, Spain, UK) approximately agree with theoretical expectations, the small number of cities in each nation and their natural variability preclude drawing strong conclusions. We demonstrate how this problem can be overcome so that cities from different urban systems can be pooled together to construct larger datasets. This leads to a simple statistical procedure to identify urban scaling relations, which then clearly emerge as a property of European cities. We compare the predictions of urban scaling to Zipf's law for the size distribution of cities and show that while the former holds well the latter is a poor descriptor of European cities. We conclude with scenarios for the size and properties of future pan-European megacities and their implications for the economic productivity, technological sophistication and regional inequalities of an integrated European urban system.
\end{abstract}
%\end{sciabstract}


% this is for the table of contents
%\setcounter{tocdepth}{1}
%\renewcommand\contentsname{}
%\tableofcontents
%\renewcommand\listfigurename{}
%\listoffigures
%\renewcommand\listtablename{}
%\listoftables

%\noindent{\bf \large Contents:}
%\begin{description}
%\item Supplementary Figures S1, ....
%\item Supplementary Table S1,...
%\end{description}

\newpage
%\linenumbers
\section{Introduction}

European nations are some of the oldest extant urban systems  in the world~\cite{pirenne_medieval_1980,braudel_civilization_1992,de_vries_european_2013,bairoch_cities_1991,abu-lughod_before_1991,nicholas_growth_1997,bowman_settlement_2011}.  Many contemporary European cities have centuries, if not millennia, of history, often stretching back to medieval or classical times. Over this long span of time, each european city has experienced periods of profound crisis alternating with booming development and has seen enormous demographic, economic, political and spatial transformations~\cite{hall_cities_1998}. From this rich historical perspective, we may expect each European city to be exceptional and unique, and not to conform to any particular quantitative expectation~\cite{hall_cities_1998,mumford_city_1989}.

However, the opposite perspective --that all cities share certain predictable quantitative properties- has slowly emerged from empirical studies and theoretical considerations developed by a variety of disciplines, including economics~\cite{henderson_urban_1991,fujita_spatial_2001,glaeser_triumph_2012}, geography~\cite{isard_location_1972,batty_size_2008,storper_keys_2013}, engineering~\cite{kennedy_evolution_2011} and complex systems~\cite{bettencourt_growth_2007,batty_size_2008,bettencourt_hypothesis_2013,bettencourt_origins_2013}.  All these disciplines explain the existence and development of cities as the result of the interplay between centripetal and centrifugal "forces", which in turn result from socioeconomic advantages of concentrating human populations in space and account for associated costs.  These are known as  {\it agglomeration} or {\it scaling} effects and constitute the foundational concepts for explaining the formation and persistence of cities anywhere~\cite{fujita_spatial_2001,duranton_chapter_2004,bettencourt_origins_2013}. Urban agglomeration effects are based on the observation of systematic changes in average socioeconomic performance, land use patterns and infrastructure characteristics of all cities as functions of city size.  Such relations are known across the sciences as {\it scaling relations}~\cite{barenblatt_scaling_1996}, which relate macroscopic properties of a system--here a city--to its scale (size). For this reason, the systematic study of such relationships in cities is known as urban scaling.

Clearly, these two perspectives -- emphasizing what is particular and what is general about cities -- are at odds with each other~\cite{henderson_urban_1991,bettencourt_urban_2010}. Each, on its own, is too simple to be fully correct, while both should be expected to play a role to a greater or lesser extent in explaining the observed properties of any city.  Thus, the interesting question is to what extent can the properties any city be predicted by general considerations and how to quantitatively assess the exceptionality of each place~\cite{bettencourt_urban_2010}. There is probably no better place to engage in this exercise than in Europe. Here, we tackle this tension by analyzing extensive evidence for the cities of the European Union where strong national context also plays an important role on top of city-specific factors.

The empirics and the theory of urban scaling are now mature enough that quantitative expectations for scaling relations can be formulated and measured in many urban systems around the world.   However, the properties of contemporary European urban systems have been studied less than those of other nations, especially the United States~\cite{pred_city_1977,henderson_urban_1991,glaeser_triumph_2012,storper_keys_2013}.  Given the movement in Europe towards greater political and economic integration, especially within the framework of the European Union, it is particularly interesting to compare and contrast persistent regional differences and continental convergence among European cities.

Comparative quantitative studies of the properties of European cities have been hampered by a lack of data for consistently defined socioeconomic units of analysis. General theoretical considerations and empirical practice lead us to view cities as integrated socio-economic networks of interactions embedded in physical space~\cite{glaeser_triumph_2012,bettencourt_origins_2013}.  Capturing this logic when delineating urban units of analyses requires that data be collected in a consistent manner across a number of multi-dimensional criteria leading to the concept of {\it functional cities}.  The definition of functional cities, as \textit{integrated socioeconomic units}, has become the gold standard for any scientific analysis of the properties of cities and urban systems. The U.S. Census Bureau has a long-standing, and arguably the most consistent, definition of functional cities, known as \textit{Metropolitan Statistical Areas} (MSAs), dating back to the 1950s and updated annually~\footnote{For historical definitions of MSAs see \url{http://www.census.gov/population/metro/data/pastmetro.html}}. MSAs consist of a core county or counties in which lies an incorporated city (a politico-administrative entity) with a population of at least 50,000 people, plus adjacent counties having a high degree of social and economic integration with the core counties as measured through commuting ties. MSAs are in effect unified labor markets reflecting the frequent flow of goods, labor and information, which in turn is a proxy for intense socioeconomic  interactions ~\cite{glaeser_economic_1995}.

In Europe, the identification of consistent (functional) territorial units has been a goal for some time now as such definitions, and the socioeconomic characteristics of the delineated territories, play important roles in the formulation of European Union (EU) policies and the allocation of EU funds, for example, the structural funds for regional development and cohesion. Until recently, several systems of territorial units have co-existed in European statistical bureaus. Most are based on the Eurostat's \textit{Nomenclature of Territorial Units for Statistics} (NUTS) classification system~\footnote{See \url{https://en.wikipedia.org/wiki/Nomenclature_of_Territorial_Units_for_Statistics}}, with urban NUTS3 corresponding roughly to integrated territorial units that can have an urban character. For these reasons, urban NUTS3 and other definitions have been the focus of several studies of agglomeration effects in European cities, using econometric analyses~\cite{ciccone_agglomeration_2002,duranton_chapter_2004,rice_spatial_2006,patacchini_local_2008,van_raan_urban_2015}.  A unification of NUTS3 into larger functional cities has also been proposed and resulted in {\it Larger Urban Units} (LUZ) and {\it Metropolitan Areas} (MAs), used in different European Union Statistics' urban audits. However, the NUTS system borrows heavily from underlying older, country specific, territorial units and, as such, is not consistently defined across different European nations.

An effort to define functional cities in a conceptually meaningful and empirically consistent manner has been recently undertaken by the Organization for Economic Cooperation and Development (OECD), in collaboration with the EU~\cite{oecd_redefining_2012}. This has resulted in a new set of harmonized metropolitan area definitions across the European Union and other OECD nations\footnote{For a detailed discussion of the how the EU and the OECD have delineated metropolitan areas go to \url{http://www.oecd.org/gov/regional-policy/Definition-of-Functional-Urban-Areas-for-the-OECD-metropolitan-database.pdf}. For maps and all data see \url{http://measuringurban.oecd.org}}. At present, these definitions represent the most consistent attempt to define functional urban areas in Europe, making contact with those of other nations such as, for example, the US, Mexico, and Japan. The advent of this dataset presents a novel opportunity to comparatively analyze the properties of European cities as a function of their population size, for which there are a number of theoretical expectations and comparative empirical evidence from other urban systems~\cite{woude_urbanization_1990,bettencourt_growth_2007,bettencourt_origins_2013}. Here we take a first step in this direction, by analyzing and discussing the scaling properties of OECD-EU Metropolitan Areas (MAs) for the five largest urban systems in Western Europe, namely France, Germany, Italy, Spain, and the United Kingdom. This allows us to consider some of the properties of these national urban systems and comment on special cases and statistical uncertainties resulting from the relatively small number of large cities in each of these nations. To tackle this problem, we show how data for most cities in the European Union can be pooled together while respecting national differences in social and economic development. In this way, we test urban scaling at the continental level, thus bypassing some of the statistical difficulties of small datasets in each nation.  This procedure also allows us to characterize regional and national differences in urban population sizes and economic performance across different European urban systems and discuss such results in the context of the pan-European population size distribution of cities.

\section{Results}

\subsection{Expectations from Urban Scaling}

We start by explicitly stating the expectations and realm of applicability of urban scaling theory as a model for analyzing the empirical properties of European cities. Details of the theory and derivation of quantitative predictions for parameters are given in~\cite{bettencourt_origins_2013}. Scaling relations are naturally written in terms of scale-free functions (power laws)~\cite{barenblatt_scaling_1996}. Other proposals, using logarithms~\cite{pan_urban_2013,sim_great_2015}, also fit the data well in the regime where these functions agree analytically\footnote{Note that $Y(N)=Y_0 N^{1+\delta}=Y_0 N e^{\delta \ln N} = Y_0 N(1+\delta \ln N + \frac{1}{2} (\delta \ln N)^2 + ...) \simeq Y_0 N \ln N/e$ , for small $\delta \ln N$. So if $\ln N < 1/\delta$, see below, these functions are essentially the same analytically.}, but implicitly introduce a scale at which the properties of cities would have to change drastically~\cite{bettencourt_hypothesis_2013}. Thus, urban scaling proposes that any city-wide property (e.g. total GDP or urbanized area), $Y$, should be written as
\begin{eqnarray}
Y (N,t) = Y_0(t) ~ N(t)^\beta e^{\xi(t)}, 
\label{eq:scaling}
\end{eqnarray}
where $N(t)$ denotes a city's population at time $t$, $Y_0(t)$ is a baseline {\it pre-factor} common to all cities and $\beta$ is a dimensionless scaling exponent (or elasticity, in the language of economics). $Y_0(t)$ is a function of time, $t$, capturing nation-wide socioeconomic development (or decline). The exponent, $\beta$, has a special status as it is {\it assumed} to be time-independent, and as such a conserved quantity across time in any urban system. Theoretical considerations show that the exponent $\beta$ is determined by general geometric considerations~\cite{bettencourt_origins_2013}, and thus that such an assumption may be justified. Scaling analysis of ancient settlement systems lends some additional empirical support to this idea~\cite{ortman_pre-history_2014,ortman_settlement_2015}. The variable $\xi(t)$ accounts for deviations in each city from the expected (power-law) scaling relationship. As written, Eq.~\ref{eq:scaling} is {\it exact} as any deviation from the power-law function in each city is absorbed into the corresponding $\xi$. Thus, the appropriateness of any scaling function to describe urban properties is tied to the statistics of $\xi$~\cite{gomez-lievano_statistics_2012,bettencourt_hypothesis_2013}.

In most urban systems thus far analyzed empirically, it has been found that the statistics of $\xi$ are approximately Gaussian, with a mean over all cities in the system equal to zero ($\langle \xi \rangle =0$)  and a quantity-dependent variance roughly of order unity ($\sigma^2_\xi =  O(1)$). The properties of the variance remain largely unexplored and require further study. The properties of $\xi$ as an approximate Gaussian random variable with zero mean justify using the simplest fitting procedure for $Y$ vs $N$, a linear relations in logarithmic variables and minimizing ordinary least squares (OLS):
\begin{eqnarray}
\ln Y_i = \ln Y_0 +\beta \ln N_i + \xi_i,
\end{eqnarray}
so that the exponent $\beta$ is the slope of the linear regression and the prefactor, $\ln Y_0$, is its ordinate at the origin ($N=1$).  This also means that the scaling relation $Y(N)=Y_0 N^\beta$ is the {\it expectation value} of the approximately log-normally distributed stochastic variable, $Y$, for a city, given its population size, $N$, or $\langle Y\rangle|_N = Y_0 N^\beta$~\cite{gomez-lievano_statistics_2012}. 

The decomposition of any urban measurable into two components, an expected value as a function of city size (scaling relation) and a local deviation $\xi$, resolves the tension between what is general and what is particular, respectively, about each city within an urban system. Because of these properties, the values of $\xi$ have been proposed and used as population size independent urban indicators to characterize the individual {\it performance} of each city relative to each other within an urban system~\cite{bettencourt_urban_2010,alves_scale-adjusted_2015}. If the dispersion (the variance, $\sigma^2_\xi$) is larger for a given urban system or a specific quantity, then the scaling relation (average expectation) is less predictive of the properties of such cities, and vice versa.  We will see some examples of both situations below. 

Empirical analyses of the scaling relations for many urban systems have suggested that there are consistent agglomeration effects~\cite{rosenthal_chapter_2004,ciccone_agglomeration_2002,bettencourt_growth_2007,gomez-lievano_statistics_2012,bettencourt_origins_2013} across city size in many urban systems, including Germany, China, Japan, the US and Brazil. This translates into expectations for specific values of the exponents, $\beta$, for different urban quantities. To calculate the expected value of these exponents, and of other associated quantities, urban scaling theory proposes a self-consistent model of socioeconomic networks embedded in urban built space as decentralized infrastructure networks~\cite{bettencourt_origins_2013}. To achieve this, it builds on a long history of earlier quantitative models~\cite{alonso_location_1977,black_theory_1999,fujita_spatial_2001} to describe a city functionally as a spatial equilibrium whose extent is set by the balance of density dependent socioeconomic interactions (centripetal forces) and transportation costs (centrifugal forces)\cite{bettencourt_origins_2013}.  This approach emphasizes the critical importance of using a functional definition of cities for empirical examinations of urban scaling: It is only for units of analysis that embody this global spatial equilibrium that the values of $\beta$ for many different urban quantities are calculated via urban scaling theory.  For other plausible urban units, such as political or administrative cities of various kinds (e.g. municipalities, counties, etc), there are at present, to the best of our knowledge, no predictions for the corresponding values of $\beta$, which may or may not appear consistent~\cite{louf_scaling:_2014}. 

For these reasons, we consider only functional cities (metropolitan areas), which are the natural definition of cities as socioeconomic systems.  We will not repeat the derivations for the values of $\beta$ here and merely restate the expectations for the exponents associated with economic performance, innovation, the volume of built infrastructure and employment, which we will contrast to data below. We have that all socioeconomic quantities (GDP, innovation) are expected to take the same exponent~\cite{bettencourt_origins_2013}, resulting in
\begin{eqnarray}
\beta_{\rm \tiny socioeconomic} = 1+\delta, \ \beta_{\rm built-infrastructure} = 1-\delta, \ \beta_{\rm employment} = 1, \ \delta = \frac{H}{2(H+2)},
\end{eqnarray}
where $2 \geq H \geq 0$ is a fractal dimension of individual movement within the city and describes aggregate socioeconomic interaction opportunities. As $H \rightarrow 0$ individuals experience cities only from their circumscribed location, social interactions cease and the city becomes spatially segregated. As a consequence, all agglomeration effects vanish, $\delta \rightarrow 0$. In this limit, population densities or economic performance are independent of city size and indeed the advantages of urban life disappear. Thus, as $H \rightarrow 0,$ cities should cease to exist, as the forces that hold them together vanish. Conversely, as $H\rightarrow 2$, individuals use the entire space of the city, which may be appropriate to describe central areas, but not the city as a whole. As $H \geq 1$ the city is fully mixing, which can be achieved at minimal movement costs for $H=1$. Thus, $H \simeq 1$ is hypothesized to be the most likely exponent~\cite{bettencourt_origins_2013}, corresponding to the simplest scenario with
\begin{eqnarray}
\beta_{\rm socioeconomic} = \frac{7}{6}, \qquad \beta_{\rm built-infrastructure} = \frac{5}{6}, \qquad \beta_{\rm employment} = 1, \qquad \delta = \frac{1}{6}.
\end{eqnarray}

This scenario is so simple, in fact, that it can only be expected to hold very approximately: some level of spatial, social and economic segregation always exists, and so does the opportunity to visit the city more extensively, especially if more accessible transportation options are available. Nevertheless, we will use these values of $\beta$ as null models and see that these simplest expectations hold surprisingly well for modern Europe, especially in the aggregate.

\subsection{Urban Scaling Properties in Five European Nations}

We proceed by analyzing the general properties of European metropolitan areas (MAs) for the largest five urban system in Western Europe. These are some of the oldest urban systems in the world. All five urban system have long roots in history, dating back to Roman times in some cases (Italy, France, parts of Britain) and the medieval period for most (i.e., Germany), and persisting through much change and transformation~\cite{hall_cities_1998,mumford_city_1989}. The UK's urban system was the first in the world to undergo the industrial revolution with well known consequences for the growth of its cities and the change in the living conditions of its inhabitants~\cite{hall_cities_1998,mumford_city_1989}. France and Germany followed suit shortly thereafter. In addition, these five urban systems have experienced very different levels of political and economic unification, with Italy and Germany being unified relatively recently and Germany being subsequently separated into East and West at the end of World War II. Finally, over the last few decades all these nations have become integrated as part of the European Union and granted free circulation of people (citizens) and capital. For all these reasons we may expect all five different urban systems to exhibit different properties.

\begin{figure}[htbp]
   \centering
   \includegraphics[scale=0.4]{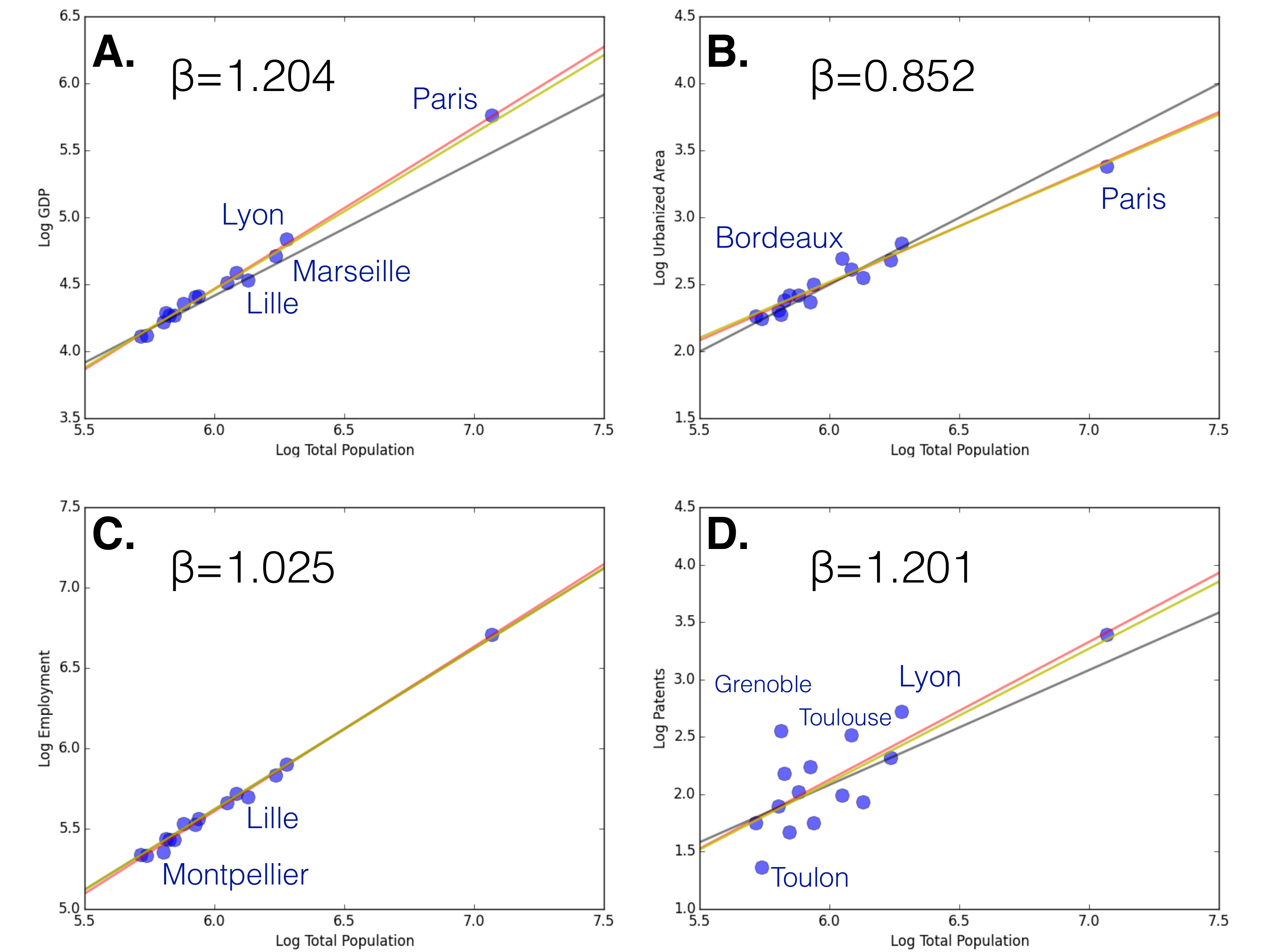} % requires the graphicx package
   \caption{The scaling of urban quantities with population size for Metropolitan Areas in France. There are 15 functional urban areas in France with populations above 500,000 people, specifically Paris, Lyon, Marseille, Toulouse, Strasbourg, Bordeaux, Nantes, Lille, Montpellier, Saint-Etienne, Rennes, Grenoble, Toulon, Nice, and Rouen. Fig 1A show the results for GDP, showing clear superlinear $\beta>1$ scaling. Lines shows the best fit (red, see Table 1, $R^2=0.98$), the simplest prediction from urban scaling theory (yellow) and a proportionality line (black) for the absence of scaling effects.  Fig. 1B shows the scaling of urbanized area ($R^2=0.92$), Fig. 1C the scaling of total Employment ($R^2=0.99$) and Fig. 1D of patents ($R^2=0.43$), as a proxy for general rates of urban innovation.  The results (red lines) are statistically indistinguishable from the predictions of urban scaling (yellow lines) within confidence intervals, but the precise value of scaling exponents is hard to ascertain because of the small sample size and the level of individual city variation.}
   \label{fig:France-scaling}
\end{figure}

\noindent {\bf France} \\
France has one of the oldest politically and economically integrated urban systems in Europe~\cite{sjoberg_preindustrial_1965,gies_life_1969}. Figure \ref{fig:France-scaling} shows the scaling behavior of all 15 cities in France with population above 500,000 people, for urban GDP, urbanized area, employment and patents.  

Urban scaling theory predicts superlinear behavior $\beta>1$ for socioeconomic quantities (GDP, patents), linear behavior for characteristics closely tied to population, such as employment ($\beta=1$), and sublinear behavior for urbanized area ($\beta<1$), expressing greater average densities in larger cities. All three trends are observed for the French urban  system with small statistical dispersion (deviations, $\xi \neq 0$), with the exception of patents, which is always a noisy quantity~\cite{bettencourt_invention_2007}, Fig.~\ref{fig:France-scaling}D.  The estimation results obtained using OLS regression on logarithmic variables produces scaling exponents that agree quantitatively with the simplest predictions from urban scaling theory. Unfortunately, the small sample size does not allow very precise exponent measurements and confidence intervals remain broad (but clearly super/sublinear as expected), especially for patents, Table~1. We will address this issue below, after seeing the problem recur for other European urban systems.

Regarding exceptions, the cities of France are exceedingly well behaved and deviations from their average scaling relation are not strong. Nevertheless, Fig. \ref{fig:France-scaling}A for GDP reveals that cities such as Marseille and Lille have smaller economies than it would have been expected for their population size. Most cities in France also have an extent of urbanized area that is very consistent with a nation-wide scaling trend; a slight exception is Bordeaux that appears larger than expected for its population. Fig.~\ref{fig:France-scaling}C shows that Lille, again, and Montpellier have lower employment than expected, a situation related to their former status as manufacturing centers. Finally, Fig.~\ref{fig:France-scaling}D expresses well known qualitative expectations that technological innovation is an important feature of French cities such as Grenoble, Toulouse and even Lyon, while Toulon, in Provence, show very little inventive activity.

\noindent {\bf United Kingdom}

\begin{figure}[htbp]
   \centering
   \includegraphics[scale=0.4]{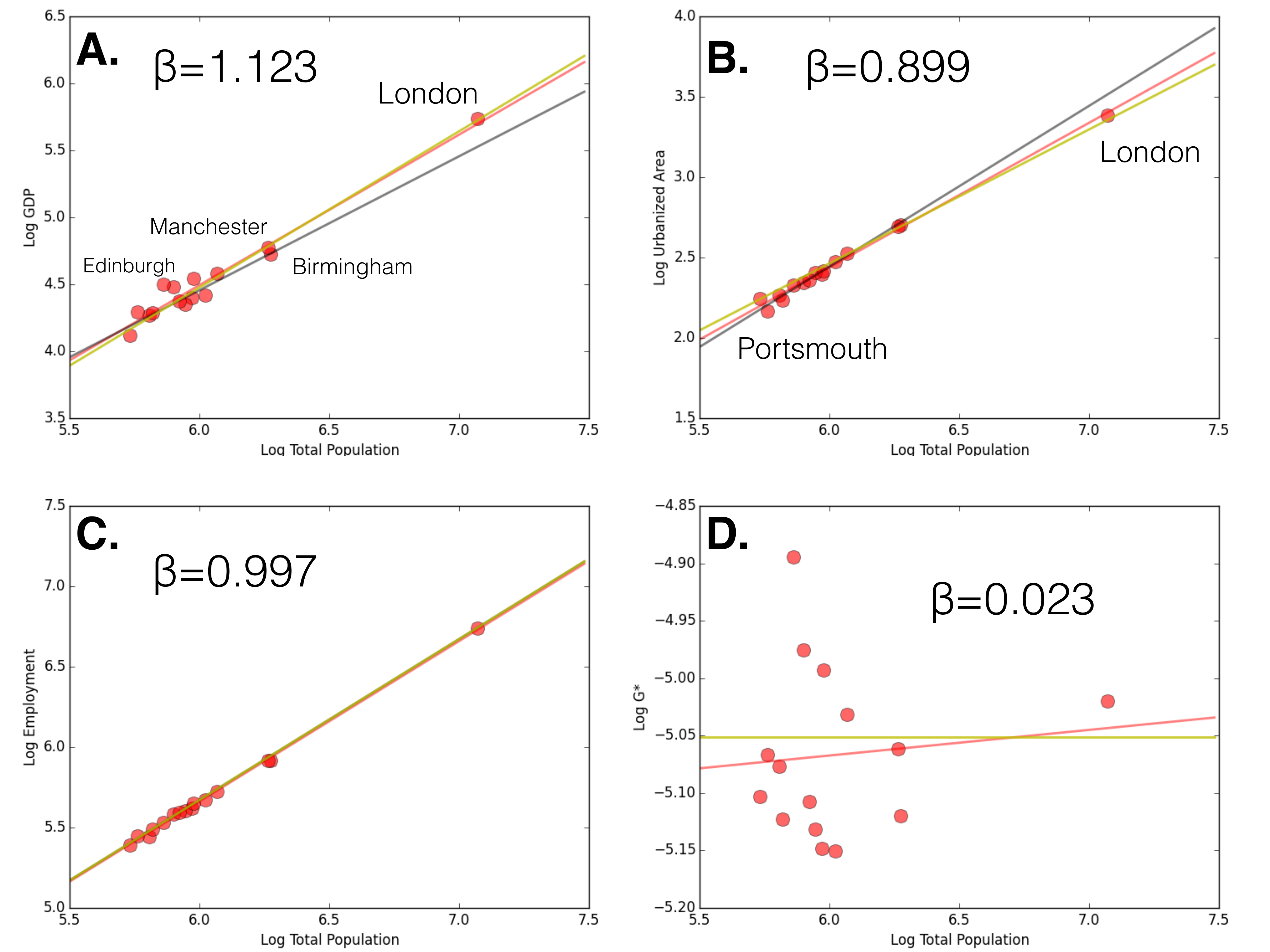} % requires the graphicx package
   \caption{The scaling of urban quantities with population size for Metropolitan Areas in the UK. There are 15 functional cities above 500,000 people in the dataset, specifically London, Birmingham, Leeds, Bradford, Liverpool, Manchester, Cardiff, Sheffield, Bristol, Newcastle, Leicester, Portsmouth, Nottingham, Glasgow, and Edinburgh. Fig 1A show the results for GDP, showing clear superlinear $\beta>1$ scaling. Lines shows the best fit (red, see Table 1,$R^2=0.92$), the simplest prediction from urban scaling theory (yellow) and a proportionality line (black) for the absence of scaling effects.  Fig. 1B shows the scaling of urbanized area ($R^2=0.98$), Fig. 1C the scaling of total Employment ($R^2=0.99$) and Fig. 1D the product of urbanized area per capita times GDP per capita ($R^2=0.00$), which is predicted by urban scaling theory to be city size invariant as observed.  We see that the best-fit results (red lines) are in broad agreement with urban scaling theory (yellow lines) and evidence from other urban systems, but that confidence intervals for parameters are wide because of the smallness of the data sample, Table~1.}
   \label{fig:UK-scaling}
\end{figure}

Analogous to France in many ways, Great Britain is also an old and fairly politically unified urban system~\cite{nicholas_later_1997}. There are also 15 MAs in Great Britain with populations above 500.000. These cities also exhibit tight scaling behavior with small deviations, Fig.~\ref{fig:UK-scaling}. Exponent estimates for GDP, urbanized area and employment agree with those of France and with the expectations of urban scaling theory, but statistically  suffer, again, from being a relatively small number of cities. Both French and British urban systems show strong macrocephaly, with Paris and London (the two largest cities in the European Union, with population 11.5M) being much larger than secondary cities in each nation. This also means that Paris and London manifest much stronger agglomeration (dis)advantages than any other cities in their national setting.

Some notable exceptions can nevertheless be identified. As is well documented, the former large scale manufacturing centers of Birmingham and Manchester (the two largest cities in Britain after London) show economies and levels of employment that are too small for their population size. Edinburgh, the political capital Scotland, behaves in the opposite direction and is richer than expected for a city of its size in the British context, Fig.~\ref{fig:UK-scaling}B. Unfortunately, the present OECD-EU data release does not provide number for patents produced in British cities. This issue has been the focus of some empirical controversy~\cite{arcaute_constructing_2014,louf_scaling:_2014}. On a separate piece,
we show that British inventors file the majority of their patents in the United States, and that when this is taken into account, strong agglomeration effects are observed. It will be important to continue to understand the nature and magnitude of measures of innovation in British cities, as there is often the perception that these activities have become too concentrated in London, but see~\cite{clifton_creative_2008}.

\noindent {\bf Spain}

\begin{figure}[htbp]
   \centering
   \includegraphics[scale=0.4]{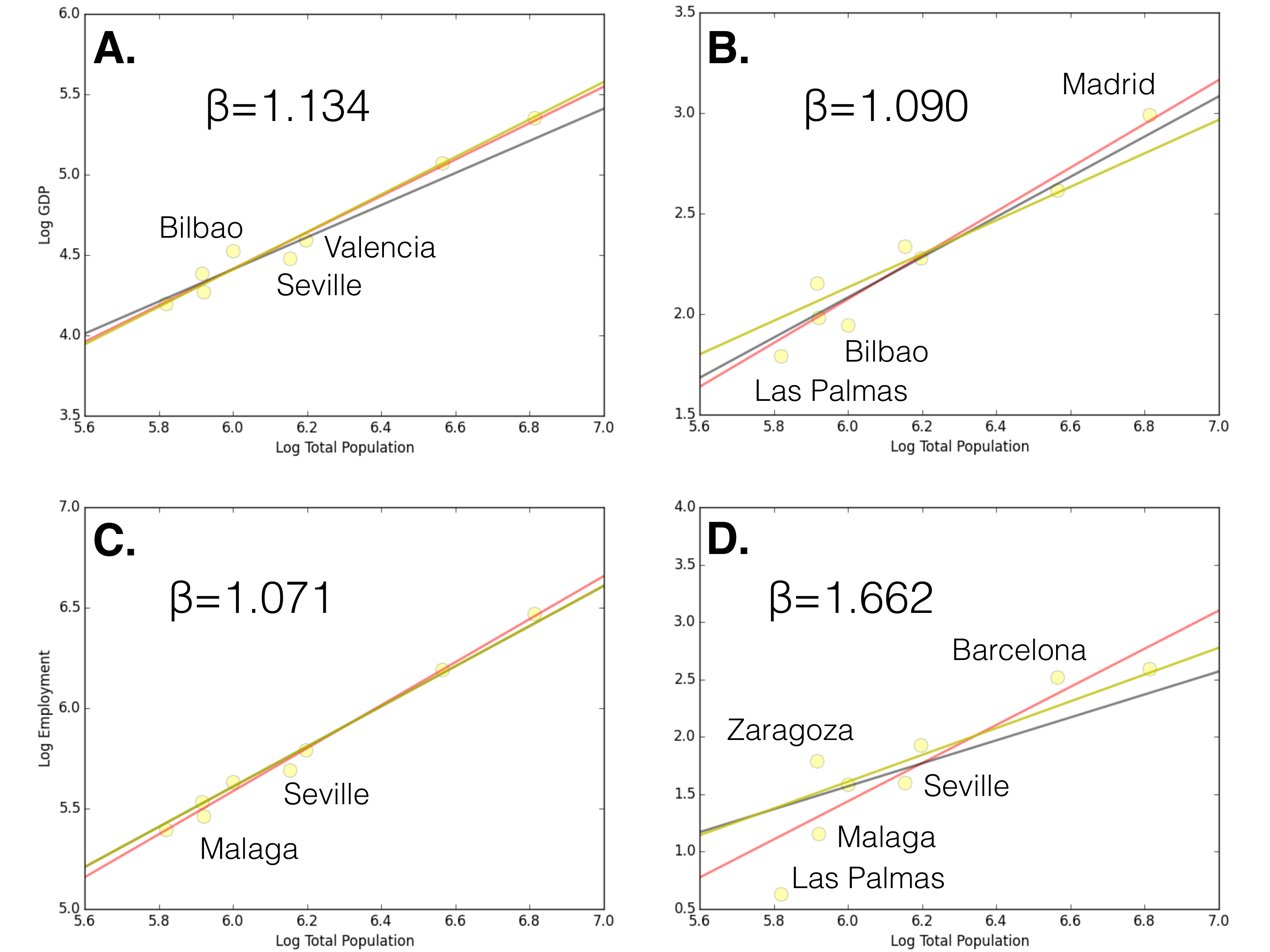} % requires the graphicx package
   \caption{The scaling of urban quantities with population size for Metropolitan Areas in Spain. There are only 8 functional cities above 500,000 people in the dataset, specifically Madrid, Barcelona, Valencia, Seville, Zaragoza, Malaga, Las Palmas, Bilbao. Fig 1A show the results for GDP, showing clear superlinear $\beta>1$ scaling. Lines shows the best fit (red, see Table 1, $R^2=0.94$), the prediction from urban scaling theory (yellow) and a simple proportionality line (black) for the absence of scaling effects.  Fig. 1B shows the scaling of urbanized area ($R^2=0.88$), Fig. 1C the scaling of total Employment ($R^2=0.98$) and Fig. 1D of patents ($R^2=0.62$), as a proxy for general rates of urban innovation. Because of the small number of large cities in Spain, as well as individual and regional variations, the confidence intervals on exponents are particularly broad, Table~1. }
   \label{Fig-Spain}
\end{figure}

Spain is the smallest of the national urban systems analyzed here, with only 8 cities above 500,000 people, see Fig~\ref{Fig-Spain}.  Nevertheless, GDP, employment and patents scale as expected, although with wider confidence intervals, see Table 1.  The urbanized area of Spanish cities appears superlinear, contrary to theory, though with a very wide confidence interval: this is largely the result of the urbanized area for Madrid, which is very large, even in the context of all large EU cities, as we shall see in greater detail below.

Among such a small number of cities the specific characteristics of particular places becomes particularly important. We see that some of the cities of Spanish Southwest, such as Seville and Malaga are poorer and less inventive than their national counterparts and that cities such a as Barcelona and Zaragoza produce a number of patents much larger than Las Palmas (Canary Islands), even when accounting for their respective population sizes.  A small urban system such as Spain's thus allows only a very general comparison with expected agglomeration effects because large uncertainties remain as to the value of average elasticities or exponents.

\noindent {\bf Italy}

\begin{figure}[htbp]
   \centering
   \includegraphics[scale=0.4]{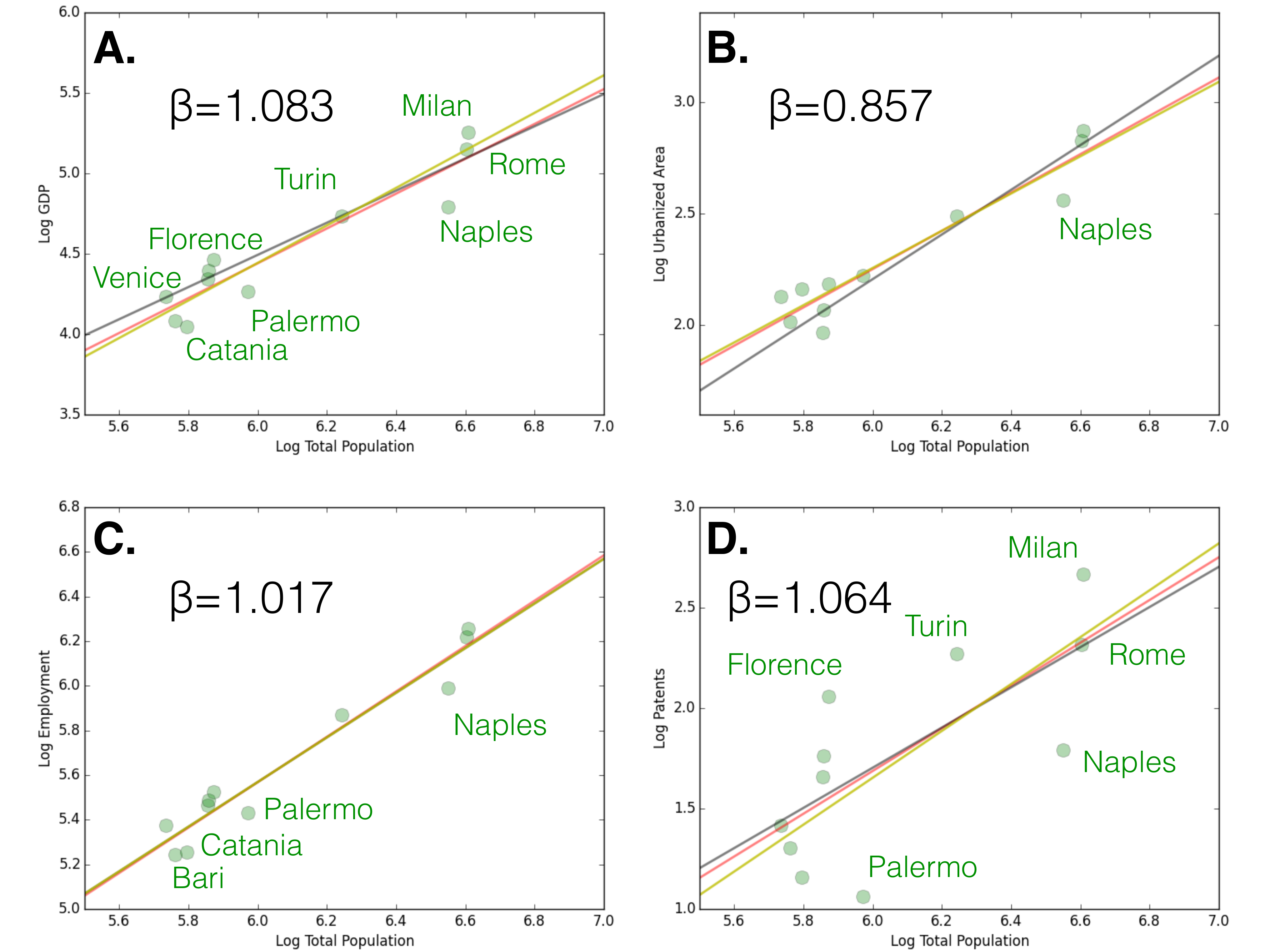} % requires the graphicx package
   \caption{The scaling of urban quantities with population size for Metropolitan Areas in Italy. There are just 11 functional cities above 500,000 people in the dataset, specifically Rome, Milan, Naples, Turin, Palermo, Genova, Florence, Bari, Bologna, Catania, Venice. Fig 1A show the results for GDP. Lines shows the best fit (red, see Table 1, $R^2=0.78$), the prediction from urban scaling theory (yellow) and a simple proportionality line (black) for the absence of agglomeration effects.  Fig. 1B shows the scaling of urbanized area ($R^2=0.83$), Fig. 1C the scaling of total Employment ($R^2=0.89$)  and Fig. 1D of patents ($R^2=0.29$), as a proxy for general rates of urban innovation. The small number of large cities in Italy, compounded by the strong North-South divide in development makes the Italian urban system far from regular from the point of view of urban scaling, resulting in lower $R^2$'s than for other nations and in broad confidence intervals for estimated exponents, Table~1.}
   \label{fig:Fig-Italy}
\end{figure}

Much like Spain, but slightly larger, the urban system of Italy is comprised of 11 cities over 500,000 people, Fig.~\ref{fig:Fig-Italy}. The most striking feature of the Italian urban system are the differences between the Northern and Southern regions of the country.  Results agree generally with the predictions of urban scaling theory but superlinear effects of GDP and patenting are a little lower than expectations, although with very wide 95\% confidence intervals. This is partly because Naples is a strong outlier along a number of dimensions: it has a small GDP, employment and number of patents for its population size and is also small in terms of its urbanized area. Moreover, Naples is not alone and other Southern Italian cities - such as  Palermo, Catania an Bari -- also underperform in term of GDP, levels of employment and  patenting.

\noindent {\bf Germany}

\begin{figure}[htbp]
   \centering
   \includegraphics[scale=0.4]{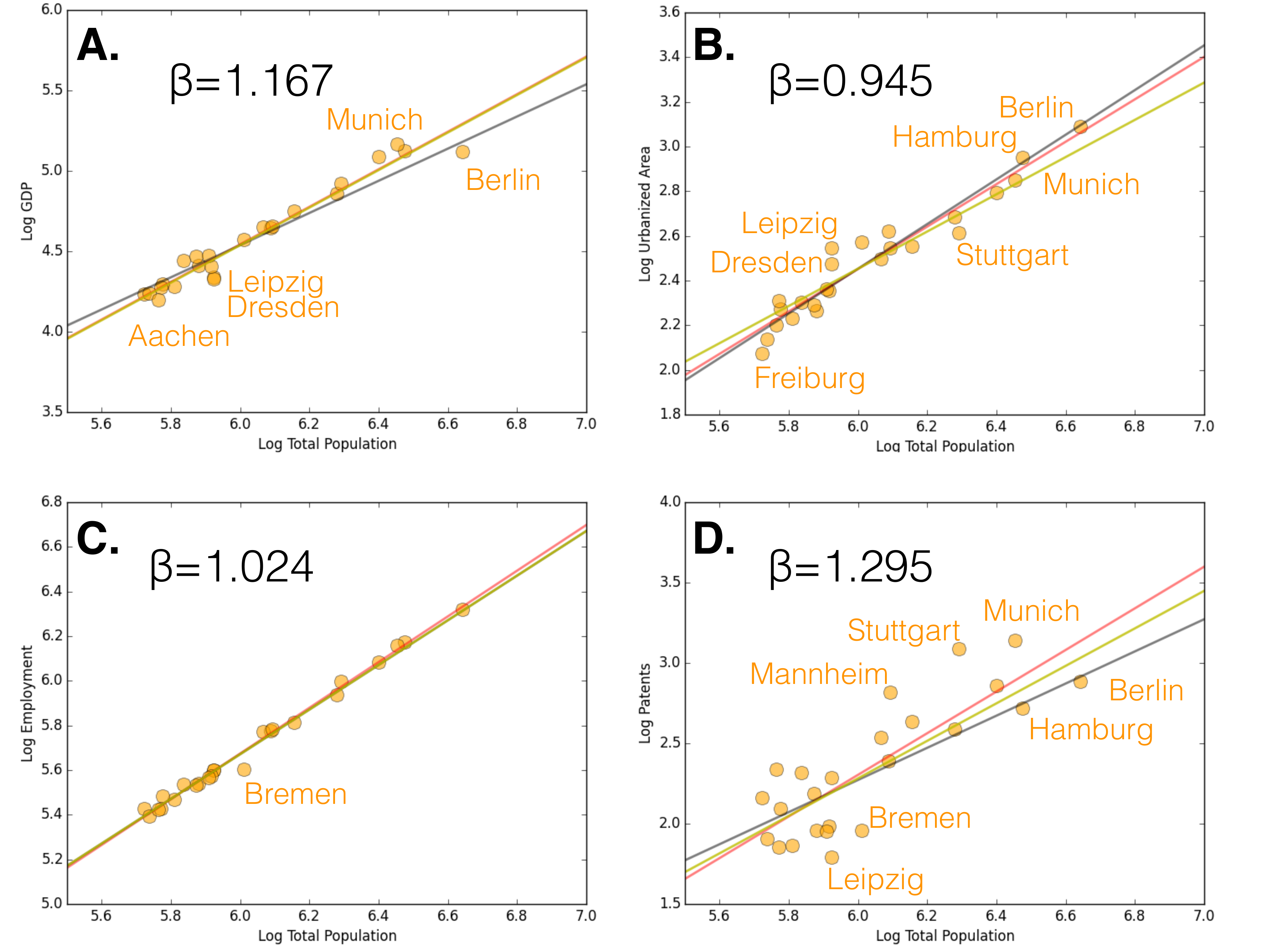} % requires the graphicx package
   \caption{Urban Scaling for Metropolitan Areas in Germany. There are 24 functional cities above 500,000 people in the dataset, specifically Berlin, Hamburg, Munich, Cologne, Frankfurt, Stuttgart, Essen, Leipzig, Dresden, Dortmund, D\"usseldorf, Bremen, Hanover, Nuremberg, Bochum, Freiburg im Breisgau, Augsburg, Bonn, Karlsruhe, Saarbr\"ucken, Duisburg, Mannheim, M\"unster, and Aachen. Fig 1A show the results for GDP, showing clear superlinear $\beta>1$ scaling. Lines shows the best fit (red, see Table 1, $R^2=0.91$), the prediction from urban scaling theory (yellow) and a simple proportionality line (black) for the absence of agglomeration effects.  Fig. 1B shows the scaling of urbanized area ($R^2=0.86$), Fig. 1C the scaling of total Employment ($R^2=0.98$) and Fig. 1D of patents ($R^2=0.47$), as a proxy for general rates of urban innovation. Despite the larger size of the German urban system, clear differences between East-West are still visible, especially in the behavior of Berlin.}
   \label{Fig-Germany}
\end{figure}

Finally, we turn to the scaling analysis of the German urban system, the largest of the five European nations analyzed here with 24 urban areas of more than 500,000 people. Fig.~\ref{Fig-Germany}A  shows the scaling of GDP for German Metropolitan Areas versus their population size. The best fit line agrees perfectly with the simplest prediction of urban scaling theory, Table~1, although the East-West divide between the nation is also apparent, with Berlin, and to a lesser extent Dresden and Leipzig standing out below the scaling line.  Fig.~\ref{Fig-Germany}B show the scaling of urbanized area, with best fit line sublinear but a little higher than urban scaling predicts. It is important to note that the East-West divide is also visible here with Eastern cities showing larger urbanized areas than expected.  Employment shows a very predictable linear trend, with the single exception of Bremen, which shows smaller number of jobs than expected, a well documented phenomenon after the decline of its shipyards and other related industries~\cite{ploger_bremen_2007}. Finally, the trend for patents is quite noisy, but shows well-known technology centers such as Munich, Stuttgart and Mannheim as strong positive outliers, while Leipzig and Bremen appear as cities with low rates of invention.

It is noteworthy that, despite strong regional and city-specific differences,  the larger number of cities in Germany allows us to start establishing the superlinear character of GDP and patenting, the strictly linear behavior of employment and the sublinear nature of urbanized area with a little more statistical confidence. The analysis of these five largest western European urban systems shows, however, that given the typical statistical dispersion in the character of each city, larger samples would be necessary to establish the actual values of scaling exponents with sufficient confidence that some testing of theory can be performed.  Such a test requires therefore a larger number of European cities, an issue that we address in the next section.

\begin{table}[ht]
\caption{Summary Scaling Exponents for GDP, Urbanized Area, Employment and Patents for European Metropolitan Areas versus population, see text and Figs. 1-7 for additional details. Square brackets show 95\% confidence intervals on exponents.}
\begin{center}
\small
\begin{tabular}{c|c|c|c|c|c|}
Nation & $N_c$ & GDP & Urbanized Area & Employment & Patents  \\
\hline
France & 15 &1.20 [1.15,1.26]&  0.85 [0.75,0.95] & 1.03 [0.98,1.07] & 1.20 [0.72,1.69]   \\
UK & 15 & 1.12 [1.00,1.25] & 0.90 [0.85,0.95] & 1.00 [0.97,1.02] &  NA \\
Spain & 8 & 1.13 [0.97,1.30] & 1.09 [0.86,1.32] & 1.07 [0.99,1.16] & 1.66 [0.95,2.37] \\
Italy & 11 & 1.08 [0.82,1.35] & 0.86 [0.68,1.04] & 1.02 [0.85,1.18] & 1.07 [0.40,1.73] \\
Germany & 24 & 1.17 [1.06,1.28] & 0.95 [0.84,1.06] & 1.02 [0.98,1.07] & 1.30 [0.92,1.67] \\
\hline
Europe & 102 & 1.17 [1.11,1.22] & 0.93 [0.88,0.98] & 1.02 [1.00,1.05] & 1.13 [0.91,1.34] \\
\end{tabular}
\end{center}
\label{default}
\end{table}%

\subsection{The Pan-European Urban System}

Estimating scaling parameters for relatively small urban systems, with less than a few dozen large cities, is fraught with procedural difficulties and typically leads to large error bands. This makes it difficult to assess the consistency of scaling parameters across nations and over time, and permits only relatively weak conclusions. Fortunately, the simple mathematical form of scaling relations provides us with a general method to pool data from across urban systems that we expect {\it a priori} have different baseline quantities, such as greater/smaller wealth in Germany/Spain.  Besides being mathematically and econometrically justified, pooling the data from the various national systems is also conceptually interesting, given the efforts at European integration via trade and financial networks, and commonalities of legal, institutional and technological frameworks. This procedure will allow us to discuss the extent of this integration along several independent dimensions.

To see this, consider the general form of scaling relation as a power law. After taking logarithms, we obtain
\begin{eqnarray}
\ln Y_i  = \ln Y_0 + \beta \ln N_i + \xi_i .
\end{eqnarray}

The average of $\ln Y_i$ over all cities, $\langle \ln Y \rangle$, is
\begin{eqnarray}
\langle \ln Y \rangle = \frac{1}{N_c} \sum_{i=1}^{N_c} \ln Y_i =  \ln Y_0 + \beta \langle \ln N \rangle,
\end{eqnarray}
where $N_c$ is the number of cities in a given urban system (nation) and where we have used the fact that $\langle \xi \rangle =0$ for a well posed fit.  

Subtracting Eq.~6 from Eq.~5, we obtain
\begin{eqnarray}
\Delta \ln Y_i = \beta \Delta \ln N_i + \xi_i,
\end{eqnarray}

with and $\Delta \ln N_i = \ln N_i - \langle \ln N \rangle$.  This relationship is now a {\it centered} scaling relation. In logarithmic scales, it is a straight line with slope $\beta$ and coordinate at the origin pinned to zero. Two different urban systems after centering share the same origin (0,0) in logarithmic axes and can be superposed. Thus, the centered scaling relation is a one-parameter model that can be used to determine the scaling exponent in a way that excludes co-variations of the intercept and exponent during estimation. 

Using this procedure we can center variables from different urban systems onto the same dataset and perform a global scaling analysis to estimate the overall scaling exponent $\beta$. Figure~\ref{Fig-All_Europe} show the result of this procedure using OECD-EU MAs for 12 European nations (102 cities). This enlarged set of observations include, in addition to the urban systems analyzed above, cities from Austria, Belgium, the Czech Republic, Netherlands, Poland, Sweden and Switzerland. We excluded urban system with 2 or fewer metropolitan areas, such as Portugal or Norway.

Fig.~\ref{Fig-All_Europe}A shows the scaling relation for GDP across Europe.  We find perfect agreement between urban scaling theory and the data using a one-parameter best fit. For urbanized area the fit diverges somewhat, but the prediction of urban scaling theory ($\beta=5/6$, no fit) hits precisely the area of both smaller and largest cities (Paris and London), which the fit misses. Agreements for employment and patents are also excellent. 

We conclude that the pooled dataset for Europe shows good general agreement with urban scaling theory and that the variability observed at the national level is a consequence of small datasets and of the levels of typical variation in cities. In this way, we explicitly see how urban scaling is an emergent property of functional cities that becomes visible statistically as more cities are considered.

\subsection{A European City with 50 million people? City size distributions and scaling}

\begin{figure}[htbp]
   \centering
   \includegraphics[scale=0.4]{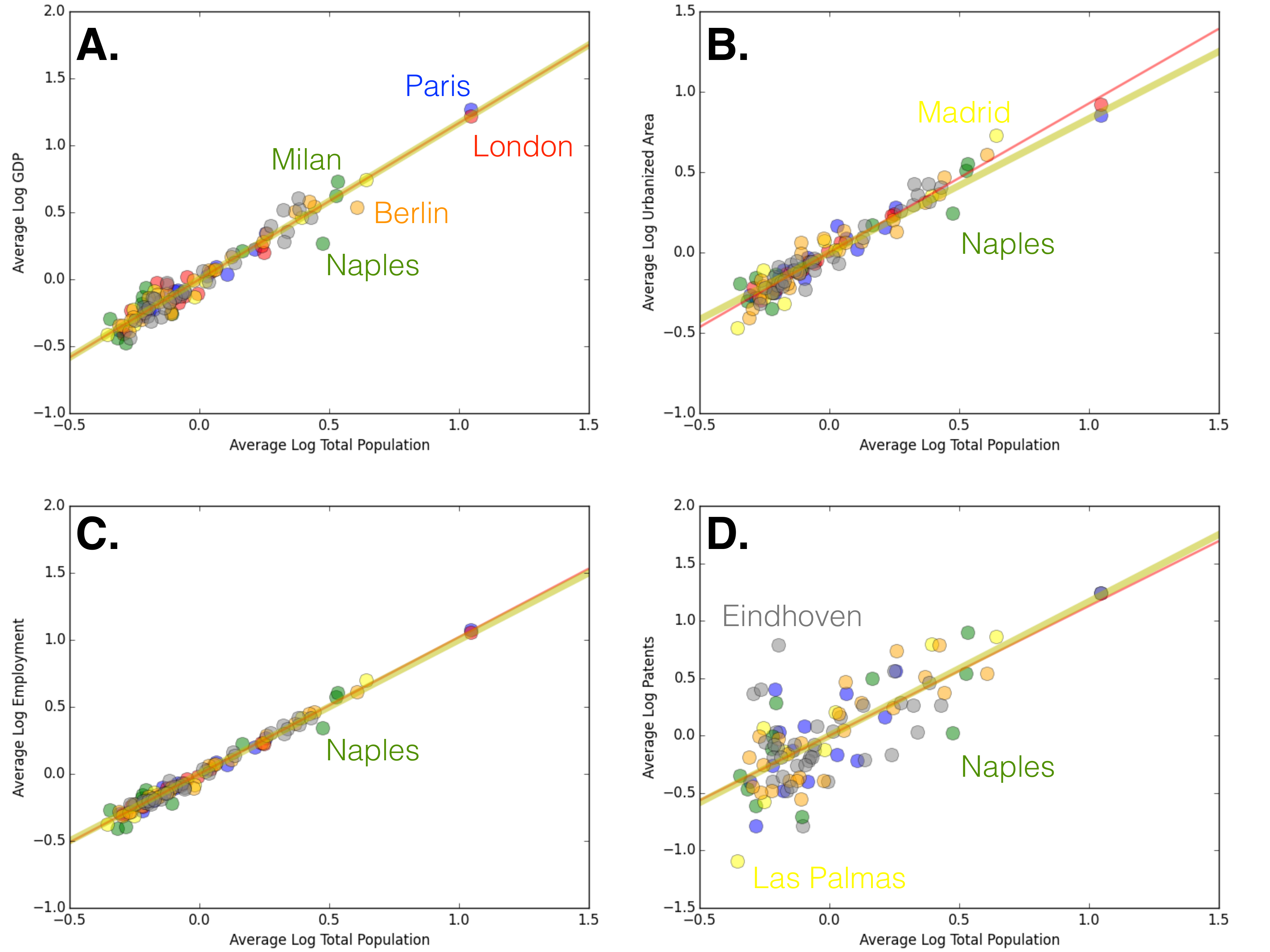} % requires the graphicx package
   \caption{The scaling of urban quantities with population size for Metropolitan Areas in Europe. These data include all urban systems in the EU and Switzerland with more than two cities above 500,000 people. This amounts to 102 functional cities in 12 nations: Austria, Belgium, the Czech Republic, France, Germany, Italy, Netherlands, Poland, Spain, Sweden, Switzerland and the United Kingdom. Because the data have been centered in each nation (see text) the expected scaling relation has intercept zero in a logarithmic plot. We have taken the simplest prediction of urban scaling theory, that the superlinear scaling exponent for socioeconomic quantities (GDP and patents) is $\beta=7/6$, that for infrastructural networks and built space (urbanized area) the sublinear exponent is $\beta=5/6$, and that for individual needs (employment) it is linear $\beta=1$ (yellow lines). With these choices there are no free parameters and a direct test of urban scaling can be performed without statistical uncertainties arising from best fits in small data samples. Fig.~\ref{Fig-All_Europe}A show the results for GDP and a nearly exact agreement with theory (best fit, $R^2=0.90$), Table~1. Fig. \ref{Fig-All_Europe}B show the results for urbanized area, the best fit gives a slightly larger $\beta$ than predicted by theory (red line, $R^2=0.88$) but fails to describe London and Paris. Urban scaling (yellow line) fits most of the data well and correctly predicts the urbanized area of London and Paris.  Employment, Fig. \ref{Fig-All_Europe}C, is also linear as expected ($R^2=0.97$). Finally, patents in Fig. \ref{Fig-All_Europe}D are noisier as this quantity is quite variable across cities and nations, but the best fit (red line, $R^2=0.30$) and prediction from urban scaling theory (yellow line) are statistically consistent and, in particular, predict well innovation rates for London and Paris.  Data for patents is not provided for the UK and is inconsistent for Poland, so that these two nations were excluded from the analysis of Fig. 1D.
}
   \label{Fig-All_Europe}
\end{figure}

\begin{figure}[htbp]
   \centering
\includegraphics[scale=0.5]{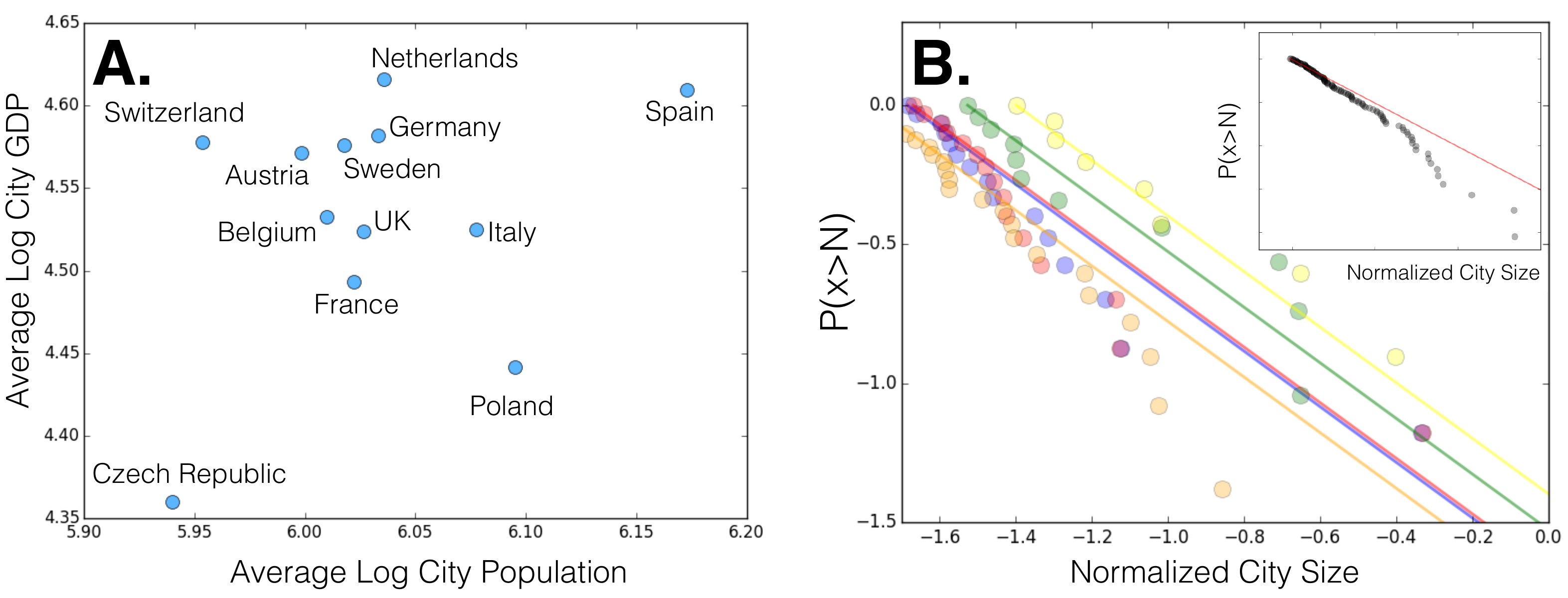} % requires the graphicx package
   \caption{The average logarithm GDP and Population size of cities in different national urban systems in Europe and their city size distributions.  A.  The average logarithmic GDP and city size across cities in each nation. These are the values used to center the data in Fig.~\ref{Fig-All_Europe}. We see no correlation, as expected, but observe that different nations are characterized by different average city sizes and GDP magnitudes, see text for discussion. B. The distribution of relative city sizes, $P(x\geq N)$, for France (blue), Germany (orange), Italy (green), Spain (yellow) and the UK (red), and for the EU, inset (112 cities $\geq$ 500,000 people).  We observe that none of the largest Western European Urban systems follows Zipf's law (solid lines): Germany lacks large enough cities; France and the UK are remarkably similar and are characterized by one very large city (Paris, London) and a set of secondary cities that are too small. When all the cities in the dataset are pooled together we find (inset) that Zipf's law (red line) is a poor descriptor of the size distribution of city sizes across the EU in the sense that Europe lacks large enough cities. This issue is explored in section 2.4, where we derive the expected size of the largest city in Europe from Zipf's law (58 million!) and predict its GDP, land area and patent production from urban scaling.
}
\label{Fig-Zipf}
\end{figure}

We have just seen how data for different urban systems can be pooled together, after centering, to provide a larger sample for which urban scaling effects can be empirically tested in a very simple and robust way. In doing this, we normalized the data for each country by the average logarithmic city size and indicator magnitude within the sample. Analyzing the magnitude of these variables for each nation gives us a sense of their convergence or divergence within the European system in terms of demographic characteristics and economic performance. Figure \ref{Fig-Zipf}A shows the average logarithmic GDP and population for cities in the 12 European nations pooled in Fig.~\ref{Fig-All_Europe}.

As expected by construction, these points show no correlation between the two variables and thus appear fairly scattered. A group of nations clusters together in the center, including Austria, Belgium, France, Germany, Sweden and the UK. These nations and their urban systems appear more integrated than other outliers, such as Switzerland, the Czech Republic, Poland or even Spain.  Tracing a vertical line of approximate same average logarithmic population, we cross, from bottom to top, France, the UK, Germany and the Netherlands. This means that with the same average city size (but different distributions, as we discuss below), these nations have cities with increasingly larger economies. In other words, the economy of the Netherlands uses its urbanization much more efficiently to produce economic value than Germany's, followed by the UK and then by France.

An analogous argument can be developed along a horizontal line, tracing nations with the same average economic performance per city but with different average city sizes. From left to right, we see, roughly along the same horizontal line, Switzerland, Austria, Sweden, Germany, and Spain. A slightly lower (poorer) horizontal line may include Belgium, the UK and Italy. This shows that Switzerland requires smaller cities to achieve the same economic performance of, say, Germany, and that Spain is able to belong to this club by having larger cities, that is, by further exploring the economic magnification effects of superlinear scaling.

Thus, the path to a richer nation overall depends on two important but uncorrelated dynamics: baseline productivity per person in cities and city sizes. Nations with lower productivity can nevertheless become wealthy as a whole by growing their cities larger, currently a worldwide phenomenon~\cite{bloom_urbanization_2008}, whereas nations with high productivity can be rich even while having relatively small cities~\cite{McKinsey_urban_2015}. Spain or Poland already have large cities but could do well to increase their baseline productivity; Switzerland or the Netherlands could become even richer simply by growing their cities further.

This brings us to the issue of city size distributions for different European nations.  This is usually summarized by Zipf's law (or rank-size rule) for the size distribution of cities ~\cite{zipf_human_2012,gabaix_zipfs_1999}. Fig. ~\ref{Fig-Zipf}B shows the counter-cumulative normalized frequency distribution, $P(N \geq x)$, for the five largest nations in the EU analyzed above. Expressed in this way Zipf's law is simply, $P(N \geq x) = {N_{\rm min}/x}$, where $N_{\rm min}$ is the smallest city size in the data set.

None of these five nations shows a city size distribution in good agreement with Zipf's law, Fig.~\ref{Fig-Zipf}.   Although there is some initial agreement for the smallest cities,  France and the UK are very similar in that they have one enormous city (Paris and London, respectively) and secondary large cities that are too small by Zipf law's expectation. Germany lacks large cities; Italy is very scattered. Spain is perhaps the urban system with the best agreement, but it is small and its two largest cities (Madrid and Barcelona) are too big compared with the Zipfian expectation.  The inset in Fig.~\ref{Fig-Zipf}B show the same distribution for all cities in Europe in the OECD-EC dataset (112 cities). We see that if we took Europe as an integrated urban system then its large cities are all too small and the Zipfian expectation fails to work at the pan-European level either. We conclude that Zipf's law, one of the oldest empirical regularities for cities, fails to give us any consistent expectation for the "right' sizes of European cities, either at the national level or in the aggregate.

The disagreement between Zipf's law for the UK (or France) could lead us to conclude, for example, that London (Paris) is too large, too expensive and too destabilizing of other cities in the UK (France), a phenomenon that policy should actively address~\cite{champion_great_2013,champion_how_2014}. However, urban scaling tells us instead that London and Paris are not at all ''anomalous". Shrinking London (or Paris), by moving population to other smaller cities, would make the UK (or France) poorer as a whole because London is the main way in which the nation explores the economic multiplicative effects of urban agglomeration. This is probably unacceptable and anathema to the spirit of the policies developed to improve the UK's urban system. Other scenarios that would make the UK (France) richer, would need to rely instead on moving population {\it up the urban hierarchy}, from smaller towns to mid-sized cities. Another, harder path, would attempt to replicate the Swiss or Dutch model and create a larger baseline productivity that allows all cities to do better without requiring demographic growth. For Germany, the path for larger wealth may have to do with growing its big cities larger, which, at least in the case of Berlin, remains a work in progress after the events of the 20th century that curbed its early explosive growth.

We finish this section with a set of conjectures for the implications of Zipf's law for Europe as a whole. According to Zipf's law, and the empirical baselines of Figure~\ref{Fig-Zipf} (inset), we can calculate that the largest city in the EU should have a population of 58 million, the second 29 million, the third about 19 million and so on. The fifth largest city would be roughly the size of today's Paris or London, with 11.5M people. Using urban scaling, we can compute the GDP, urbanized area, patent production and many more quantities~\cite{bettencourt_origins_2013} for these hypothetical cities. The GDP of such largest city would be 6.5 times larger than that of Paris or London today, which corresponds to an increase of 30\% per capita in these cities. A similar increase in density in these cities (decrease in urbanized area per capita) would also be expected, but note that this would result in a density for Paris of about 26,000 people/km$^2$, still much lower than earlier 20th century densities and than some of the most exciting parts of Paris (presently, the 11$^{\rm th}$ {\it arrondissement} is the densest in Paris with a population density of 40,000 people/km$^2$).

Creating such cities in Europe through the vigorous growth of Paris, London, Berlin, Madrid, Milan and others to such enormous sizes may certainly have its drawbacks. Continued urban system integration at the pan-European level is all but certain to grow these population centers disproportionately, further increasing inequalities between richer and poorer regions in Europe. But, at the same time, it would create a truly international culture in Europe, beyond today's heritage of older nationalisms and unleash massive technological and economic growth of the kind most European nations can currently only dream of.

\section{Discussion}

We have analyzed the scaling properties of European Metropolitan Areas, defined by the most recent joint effort by the OECD-EU to create harmonized functional cities~\cite{oecd_redefining_2012}. Arguably, these functional urban units constitute the best consistent definition of socioeconomic cities in Europe constructed to date and allow for improved comparative analyses of urban properties throughout Europe and beyond. Using these spatial units, we find support for the quantitative predictions of urban scaling theory regarding scaling exponents, especially for France and the UK and in the aggregate of EU nations. This shows that urban scaling with the specific elasticities (exponents) discussed here is exhibited by urban systems whose constituent units are indeed functional urban units, even in Europe. Other plausible urban units of analysis (such as political cities) are likely poorer approximations to cities as socioeconomic units and should be expected to exhibit different elasticities or, possibly, no clear scaling at all~\cite{louf_scaling:_2014}.  By using units of analysis whose spatial delineations actually capture urban functionality, the British urban system and especially France, are shown to exhibit expected scaling behavior, despite many historical and contemporary peculiarities~\cite{arcaute_constructing_2014}. There is also broad empirical agreement between the scaling patterns of European MAs and the properties of other large urban systems, such as the United States, China or Brazil~\cite{bettencourt_growth_2007,bettencourt_hypothesis_2013}.  The current OECD-EU dataset covers urban areas with populations larger than 500,000 people so it will be interesting to explore in future analyses smaller harmonized functional cities. Given the density and compactness of many European regions these spatial units may in many cases be difficult to define unambiguously.

Despite the effort and the rationale involved in this most recent definition of functional cities in Europe, it should be expected that such definitions will continue to be improved in the future and we look forward to revisiting our empirical findings at such times. Urban scaling analysis provides a general simple expectation for many of the properties of a city in relation to its urban system and, as a consequence, it constitutes a means to identify places with exceptional properties, good and bad. Such deviations can be the result of true local exceptionality or of data issues. Our analysis flags a number of European MAs as exceptional, in their regional context and in Europe at large. The strongest deviations from scaling for GDP are observed for Naples, Italy and for Berlin, Germany. In both cases these cities are either too large in population for their economic performance, or have economies that are too small for their populations. Berlin, and other former East Germany cities, having endured the ravages of WWII and the cold war, still tend to underperform compared to their their West German counterparts. But Berlin, given its history and recent policy interventions, clearly stands on its own as a particular case. It would be interesting to analyze the sensitivity of the economic performance for these cities versus their spatial definitions further and to follow their temporal evolution closely. Similarly, and possibly related, the urbanized area of Naples appears too small and that of Madrid, Spain too large. In this way, urban scaling analysis, can be used to point the way towards better quantitative and systematic understanding of the exceptionality of specific places and for better understanding their specific contingencies and histories from a quantitative perspective. 

On the strength of the results presented here and those from previous studies of other contemporary and older urban systems, we can conclude that urban scaling is a stronger statistical regularity for functional cities than the rank-size distribution of city size, also known as Zipf's law. Indeed different European nations show very different rank-size distributions, with e.g. France or the UK showing strong primacy (as Paris and London are 4-5 times larger than the second largest city) whereas in Germany, Spain, or Italy the opposite is true and several large cities co-exist. Despite these well known and very variable patterns in the size distribution of cities, agglomeration effects and the resulting scaling relations persist with greater regularity in each system and especially in the analysis that pools all nations in Europe together. From the perspective of urban scaling, London or Paris are not exceptional cities at all. Rather, their properties are just what one should expect for a British or French city of their population size. Thus, it will be interesting to continue to develop urban theory that brings together the theoretical insights behind scaling and agglomeration effects with those that predict the size distribution of cities.

Today, Europe remains a less urbanized continent than North America or developed Asia, with an overall urbanization rate of about 70\%.  Paris and London are     growing slowly, with annual growth rates of 0.68\% and 1.18\%, respectively, typical of most EU cities. Madrid is Western Europe's fastest growing large city with an annual growth rate of 1.8\%~\cite{oecd_redefining_2012}.  At this pace, London would double its population in 61 years, Paris in 106, and Madrid in 40 years. As a thought experiment, we extrapolated the expectations of Zipf's law for the size of the largest cities in the European Union to predict a city with population above 50M people and a number of other very large pan-european megacities. Using urban scaling theory, we predicted (conservatively, in the absence of additional economic growth) that such a city would be an economic colossus, with a GDP and invention rate {\it per capita} 30\% larger than those of Paris and London today. Thus, the rise of pan-european megacities would create tremendous magnification effects to wealth creation and technological invention that would keep Europe on par with other large and fast developing nations, such as the US, Japan, and future developed versions of China and India. Such massive transformations of Europe's urban system would, however, also severely exacerbate regional inequalities by further amplifying the wealth, technology and organizational sophistication of the richest areas of Europe today.

\section*{Acknowledgements}
We thank Deborah Strumsky for discussions and Daniel Sanchez Serra for clarifications of the OECD-EU definitions of harmonized functional cities. This work was partially funded by the MacArthur Foundation (grant 13-105749-000-USP) and the Army Research Office Minerva program (Grant W911NF1210097).  We also thank the Arizona State University/Santa Fe Institute Center for Biosocial Complex Systems for support.

\bibliographystyle{naturemag}

\bibliography{urban_scaling_europe}
%\section*{Figure Captions}

\end{document}